\documentclass[nonacm]{acmart}

\usepackage{hyperref}
\usepackage{subfig}
\usepackage{dialogue}
\usepackage{xcolor}
\usepackage{soul}

\AtBeginDocument{%
  \providecommand\BibTeX{{%
    \normalfont B\kern-0.5em{\scshape i\kern-0.25em b}\kern-0.8em\TeX}}}

\begin{document}

\title{"Is It My Turn?" Assessing Teamwork and Taskwork in Collaborative Immersive Analytics}

 \author{Michaela Benk}
 \authornote{Both authors contributed equally to this research.}
 \email{mbenk@ethz.ch}
 \orcid{0000-0002-8171-320X}

 \author{Raphael P. Weibel}
 \authornotemark[1]
 \email{raweibel@ethz.ch}
 \orcid{0000-0002-8854-7507}
 \affiliation{
   \institution{Mobiliar Lab for Analytics, ETH Zurich}
   \city{Zurich}
  \country{Switzerland}
}

 \author{Stefan Feuerriegel}
 \orcid{0000-0001-7856-8729}
 \affiliation{
   \institution{LMU Munich}
   \city{Munich}
   \country{Germany}}

\author{Andrea Ferrario}
 \orcid{0000-0001-9968-9474}
 \affiliation{
   \institution{Mobiliar Lab for Analytics, ETH Zurich}
   \city{Zurich}
   \country{Switzerland}
}

\begin{abstract}
Immersive analytics has the potential to promote collaboration in machine learning (ML). This is desired due to the specific characteristics of ML modeling in practice, namely the complexity of ML, the interdisciplinary approach in industry, and the need for ML interpretability. In this work, we introduce an augmented reality-based system for collaborative immersive analytics that is designed to support ML modeling in interdisciplinary teams. We conduct a user study to examine how collaboration unfolds when users with different professional backgrounds and levels of ML knowledge interact in solving different ML tasks. Specifically, we use the pair analytics methodology and performance assessments to assess collaboration and explore their interactions with each other and the system. Based on this, we provide qualitative and quantitative results on both teamwork and taskwork during collaboration. Our results show how our system elicits sustained collaboration as measured along six distinct dimensions. We finally make recommendations how immersive systems should be designed to elicit sustained collaboration in ML modeling. 

\end{abstract}

\keywords{Augmented Reality, Immersive Analytics, Collaboration, Machine Learning, Interpretability}

\maketitle

\section{Introduction}

Immersive analytics (IA) \cite{chandler2015immersive} explores new ways to visualize and interpret data \cite{Skarbez2019ImmersiveAT}, allowing users to turn the space surrounding them into ``a canvas for data analysis'' \cite{Batch2020ThereIsNoSpoon}. IA aims to leverage human cognition and spatial stimuli to support the understanding of abstract data, offering embodied data exploration through spatial immersion, and increased engagement, among others \cite{Batch2020ThereIsNoSpoon, Billinghurst2018CollaborativeIA}. Prior works in IA have highlighted the potential of immersive systems for collaborative tasks \cite{Butscher2018ClustersTA, Ens2021GrandCI}. To this end, \textbf{collaborative immersive analytics} (CIA) \cite{Billinghurst2018CollaborativeIA} aims at understanding the use of immersive technologies in collaborative data analysis \cite{Billinghurst2018CollaborativeIA, Sereno2020CollaborativeWI, Ens2021GrandCI}. Collaborative data analysis using traditional 2D tools has been studied extensively in the past \cite{Keim08visualanalytics, Billinghurst2018CollaborativeIA}, yet little is known about how collaboration in IA unfolds. Emerging CIA research challenges include (1) defining application scenarios, (2) designing systems that also integrate traditional tools and interfaces, and (3) assessing all the dimensions of collaborative work \cite{Ens2021GrandCI}. 

In industrial applications, collaboration between different stakeholder groups (e.g., ML engineers, domain experts, and managers) is key for machine learning (ML) modeling. Hence, current ML practice in industry relies on collaboration among interdisciplinary teams \cite{HBR,HDSR,NatMachIntell, Benk-Weibel}. In addition to the complexity of ML itself, this presents challenges for scaling ML to the broader workforce \cite{HBR}.  ML activities can focus on the training and validation of the ML models used in products and services or the understanding of their predictions using ML interpretability methods, such as visual explorations of the outcomes or counterfactual explanations \cite{hopkins2021machine, Wachter2017CounterfactualEW}. These methods are key for auditing and improving models in industry, assessing risks, such as discrimination and bias, and discovering patterns in data \cite{holstein2019improving,miller2019explanation}. They can affect the levels of users' confidence and acceptance of model predictions \cite{binns2018s,Kaur2020InterpretingIU, PoursabziSangdeh2021ManipulatingAM}. 

As IA has the potential to offer novel ways to visualize and interact with data, researchers have started to highlight the need to facilitate and improve collaboration in advanced data science activities  \cite{Wang2019HowDS, Zhang2020HowDD} and to focus their research endeavors on realistic, industry-relevant IA systems \cite{Fonnet2021SurveyOI}. Although research has considered ML workflow in single-user settings \cite{Spinner2020explAInerAV}, works in the computer-supported cooperative (CSCW) domain have only recently begun to explore collaborative practices from the point of view of data science and interdisciplinary teams \cite{Zhang2020HowDD,passi2018trust,Wang2019HowDS}. Moreover, research calls for interdisciplinary work between IA and CSCW \cite{Fonnet2021SurveyOI} and industry-relevant applications to study ML modeling \cite{passi2018trust}. However, there is no evidence to support the understanding of collaboration in IA for applications that match complex, real-world collaborative tasks with users of different professional backgrounds and levels of domain knowledge, such as in the case of ML model interpretability activities \cite{ens2019revisiting}. We fill this void by offering evidence from interdisciplinary teams solving real-world ML tasks.

In this work, we examine how collaboration in IA unfolds and can be assessed when users with different professional backgrounds and levels of ML modeling knowledge interact with an IA system to solve ML tasks collaboratively. To do so, we designed a CIA system and used it in an observational user study where pairs of users followed a realistic roleplay scenario and solved ML tasks collaboratively. We assessed the taskwork and teamwork dimensions of collaboration during ML modeling. To this end, we measured users' performance and adapted a collaborative visual analytics methodology, called pair analytics \cite{AriasHernandez2011PairAnalytics}, to our scenario. This allowed us to assess collaboration without interrupting the interactions of the pairs of users and identify six dimensions of collaboration from the CIA literature \cite{Billinghurst2018CollaborativeIA}. Moreover, we evaluated user experience and system usability to assess the impact of the IA system on collaboration.

Our work is the first to understand collaboration dynamics when users solve industry-relevant ML-related tasks using a CIA system. Our results identified drivers that foster sustained collaboration among users, as reflected by assessing their collaboration dynamics. We identified key takeaways and issues that can inform future work in CIA for ML and support the design of IA systems for ML applications in industry that target interdisciplinary teams of professionals. While other works have studied collaboration in AR, our work is unique in that we address the specific characteristics in ML modeling, namely the complexity of ML tasks, the interdisciplinary setting, as well as the need for ML interpretability.

\section{Related work}

IA aims at exploring new and intuitive ways of visualizing, exploring, and understanding data with immersive and spatially-oriented technologies that may complement traditional 2D systems \cite{chandler2015immersive,Reski-et-al_2020}. IA technologies currently include virtual-, augmented-, and mixed-reality devices \cite{bach2019interaction}. In the following, we review the related works and research gaps in three areas of relevance for our study: (1) collaborative IA, (2) system design in IA, and (3) ML interpretability as an application scenario for CIA. 

\subsection{Collaboration in Immersive Analytics}

Collaborative immersive analytics (CIA) is ``the shared use of immersive interaction and display technologies by more than one person for supporting collaborative\footnote{Collaboration is ``a coordinated, synchronous activity that is the result of a continued attempt to construct and maintain a shared conception of a problem'' \cite{Roschelle1995TheCO}.} analytical reasoning and decision making'' \cite{Billinghurst2018CollaborativeIA}. As immersive technologies have the means to create collaborative spaces and visualize data in different space-time settings, CIA has the potential to facilitate and transform collaborative work \cite{ens2019revisiting}. Thus, researchers have called for studies to understand how such technologies can support effective collaboration in various scenarios \cite{Ens2021GrandCI, Skarbez2019ImmersiveAT, Billinghurst2018CollaborativeIA}.

Collaborative scenarios are commonly classified according to where they occur physically (distributed vs. co-located) and when they occur in time (synchronous vs. asynchronous) \cite{Isenberg2011CollaborativeVD, Billinghurst2018CollaborativeIA}. These scenarios can include collaborative data exploration \cite{Reski-et-al_2020}, document search and analysis \cite{Isenberg2012CoLocatedCV}, predictive modeling using computational notebooks \cite{Wang2019HowDS}, or the identification of clusters, trends, and outliers in multidimensional data \cite{Butscher2018ClustersTA, Kraus2020TheIO}. In co-located synchronous collaborations, augmented reality (AR) can offer a number of advantages over other immersive technologies, such as a greater awareness of shared artifacts, collaborators' presence, and natural communication and improved coordination \cite{Billinghurst2018CollaborativeIA, Cordeil2017ImmersiveCA, Butscher2018ClustersTA}. This, in turn, enables the forming of a shared understanding among users \cite{Heer2008DesignCF, Wang2019HowDS, Dourish1992}. 

To this end, a number of recent works focused on the use of augmented reality in CIA. For example, \citeauthor{Butscher2018ClustersTA} proposed \emph{ART}, a collaborative tool based on augmented reality, that facilitates the visualization of multi-dimensional data \cite{Butscher2018ClustersTA}. \citeauthor{Bschel2021MIRIAAM} introduced \emph{MIRIA}, a mixed-reality toolkit to support a novel type of in-situ visual analysis \cite{Bschel2021MIRIAAM}. \citeauthor{Ens2021UpliftAT} designed \emph{Uplift}, a prototype that combines several technologies, such as a central tabletop display with augmented reality, to support casual collaborative analysis of microgrid data \cite{Ens2021UpliftAT}. Finally, \citeauthor{Ferrario2020ALEEDSAAR} have recently introduced the proof of concept \emph{ALEEDSA}, \emph{A}ugmented \emph{LE}arning \emph{D}ata \emph{S}cience \emph{A}pp, to perform interactive ML with immersive AR \cite{Ferrario2020ALEEDSAAR}. 

The above works offer valuable design considerations and suggestions for the development of future CIA systems; however, they are affected by limitations. (1)~Prior studies have focused on collaboration among users with similar backgrounds, rather than on interdisciplinary teams, e.g., those including data scientists, domain experts, such as risk or legal experts, and managers, as it is common in industry applications \cite{Wang2019HowDS,hopkins2021machine,Ferrario2020ALEEDSAAR}. Considering collaboration among interdisciplinary teams is important to offer realistic evidence of real-world ML in industries \cite{HBR,HDSR}, and presents one of our contributions. (2)~Prior works offer little insight into the characteristics of collaborative interactions in immersive environments and whether or how they differ from more traditional settings. Roschelle and Teasly argue that collaboration takes place in ``a negotiated and shared conceptual space, constructed through the external mediational framework of shared language, situation, and activity'' \cite{Roschelle1995TheCO}. Therefore, to build effective CIA systems, it is necessary to focus on how conceptual spaces are constructed in immersive settings and how they differ from other technologies. (3)~The evaluation of collaboration in IA is still fraught with methodological challenges \cite{cook2005illuminating,Ens2021GrandCI}. As CIA systems are designed to support taskwork and teamwork, both users' performance and dynamics of collaboration need to be evaluated using quantitative and qualitative measures taking into account possible spatial, time, and device asymmetries among users \cite{Ens2021GrandCI,Billinghurst2018CollaborativeIA}. To this end, different authors have called for the introduction of evaluation frameworks comprising qualitative and quantitative evaluation methods for CIA scenarios \cite{Billinghurst2018CollaborativeIA,Ens2021GrandCI}. 

We study collaboration in CIA for ML modeling. To the best of our knowledge, our work is the first in which limitations (1)--(3) of existing literature are addressed.

\subsection{System design for IA}

Prior research in IA has recognized that, even though immersive technologies can potentially offer benefits over traditional systems (i.e., desktop computers) in displaying and interacting with data \cite{Munzner2014VisualizationAA,Ens2021GrandCI}, users may be comfortable with traditional tools and be hesitant to move to fully immersive environments \cite{Wang2020TowardsAU, Grubert2020BackTT, Besanon2017HybridTI}.

Therefore, different authors have proposed systems where immersive technologies are used as extensions of traditional interfaces, such as keyboard and mouse, or touch screens \cite{Wang2020TowardsAU,Reski-et-al_2020,Ferrario2020ALEEDSAAR,Grubert2020BackTT}. For example, \citeauthor{Wang2020TowardsAU} investigated the use of a setup to help domain experts (particle physicists) explore and understand 3D data \cite{Wang2020TowardsAU}. The authors used a Microsoft HoloLens head-mounted display (HMD) to extend traditional desktop views, finding that participants appreciated the hybrid approach and the interactivity of the AR application \cite{Wang2020TowardsAU}. \citeauthor{Reski-et-al_2020} explored the use of a ``hybrid'' web application using virtual reality to support asymmetrical collaboration between two users during synchronous data analysis \cite{Reski-et-al_2020}. The authors validated the approach in a user study, investigating participants' awareness, deixis, and dynamics during collaboration. As a result, such systems offer a way of complementing existing tools or transitioning from familiar interactions that usually involve a seated environment with a mouse and keyboard as the input devices, to new interactions, such as gestures, speech, and a more physical involvement, using HMDs. However, learning new gestures and interactions may be difficult for users \cite{norman2010natural}, and physical involvement may result in discomfort, such as headache, nausea, motion sickness, or fatigue  \cite{Ens2021GrandCI}. \citeauthor{Ens2021GrandCI} thus emphasized the necessity to improve systems for IA, allowing users to interact with traditional and immersive interfaces, and minimizing the negative effects stemming from their extended usage \cite{Ens2021GrandCI}. However, as opposed to previous works, we focus on collaboration and IA for ML interpretability.

\subsection{Application scenarios for CIA: Machine learning interpretability}

ML aims at designing computing systems, called models, that learn automatically from data \cite{jordan2015machine}. ML models generate predictions to support human decision-making across different domains, such as healthcare, financial services, marketing, and education \cite{jordan2015machine}. ML interpretability \cite{lipton2018mythos,miller2019explanation,Kaur2020InterpretingIU, Carvalho2019MachineLI, PoursabziSangdeh2021ManipulatingAM} refers to the scientific efforts aiming at the design of methods to validate ML models, explain their inner workings and their predictions by means, for example, of the exploration of outcome visualizations or the use of counterfactual explanations \cite{Wachter2017CounterfactualEW}. Counterfactual explanations are hypothetical scenarios in which a data point is changed to yield a different machine learning outcome \cite{Wachter2017CounterfactualEW}. For example, consider the case of a customer asking why she did not receive a loan. Then, a counterfactual explanation may state \emph{``you would have received the loan if your income was above \$10,000''} (all else being equal) \cite{mothilal2020explaining}.

Using ML interpretability, humans can effectively audit and improve their models, assess risks (e.g., discrimination and bias \cite{chouldechova2017fair,chouldechova2020snapshot}), and discover patterns in data. In industry, both ML modeling and interpretability are collective activities involving different types of professionals, such as ML engineers, managers, domain experts, and other staff  (e.g., legal and ethics experts), among others \cite{hopkins2021machine,hong2020Human}. In the CSCW domain, research around data science topics has increased rapidly \cite{Zhang2020HowDD}. Researchers have started exploring the need for collaboration and interdisciplinarity in data science use cases, by ``turning to interdisciplinary human teams to combine domain knowledge with data science knowledge'' \cite{wang2019human}, calling for the ``inclusion of diverse expertise in data science teams'' \cite{passi2018trust}, and ``including domain experts as effective users of data science tools'' \cite{wang2019human}. In particular, Passi and Jackson conducted ethnographic fieldwork with a corporate data science team comprising business experts and ML experts collaborating on different use cases where ML models are used for customer management \cite{passi2018trust}. 

Despite the potential of using 3D immersive visualization for data  \cite{chandler2015immersive,mahyar2014supporting,hackathorn2016immersive,Ferrario2020ALEEDSAAR}, the opportunities to assist users in ML interpretability for industrial applications is still in its infancy. Importantly, such applications must adhere to current ML practice in industry where ML modeling involves the collaboration among interdisciplinary teams \cite{HBR,HDSR,Wang2019HowDS}.  To the best of our knowledge, no study has conducted a systematic evaluation of collaboration in IA systems aimed at ML interpretability. 

Related but different from our work are a number of tools that have been developed to write code, inspect data, and visually inspect ML models to debug and validate them, yet outside of AR and IA \cite{Wang2019HowDS,Spinner2020explAInerAV, Arrieta2020ExplainableAI,gehrmann2019visual,biecek2018dalex,Wang2019HowDS}. As such, these tools have clear differences: (1) they are displayed on a 2D screen, (2) comprise multiple interactive functionalities via keyboard and mouse input, and/or (3) usually involve a single user without collaboration. However, ML modeling nowadays involves collaborative work \cite{Wang2019HowDS}. Moreover, collaborative tools, such as synchronized notebooks, are typically used by users with the same domain knowledge and similar professional background \cite{Wang2019HowDS} and thus are not intended to support collaboration with other stakeholders (e.g., managers, domain experts). Therefore, it remains an open challenge how to design systems for industry-relevant and realistic CIA applications that focus on ML modeling and promote collaboration among the members of interdisciplinary teams.

\section{Methodology}

\subsection{System design}
\label{sec:system_design}

Our CIA system is an upgraded version of the system originally proposed by \citeauthor{Ferrario2020ALEEDSAAR} \cite{Ferrario2020ALEEDSAAR}. Our CIA system  allows for (1) co-located (same physical space), and (2) synchronous (same time) collaboration of pairs of users performing interactive ML interpretability activities. It comprises three components: (1) an \textbf{AR application} running on HMDs, (2) a \textbf{web application} running in browsers, and (3) a \textbf{backend server}. The components are described below.

\subsubsection{AR application} 

The AR application runs on Microsoft HoloLens~2 devices. It comprises two main interfaces: (1) the model designer, and (2) the prediction explorer. The interfaces allow for visualizing 3D interactive content using the HMD devices, which we refer to as ``visualization'' for the remainder of this work.

(1)~The model designer is an interactive interface for training ML models. The training of ML models is performed by specifying variables and hyperparameters defining the model architecture (e.g., the number of hidden layers of a neural network). For this experiment, the number of available options was limited based on the ML task at hand, and only the correct selections would lead to a trained model. 

\begin{figure}[h]%
    \centering
    {{ \includegraphics[width=0.474\linewidth]{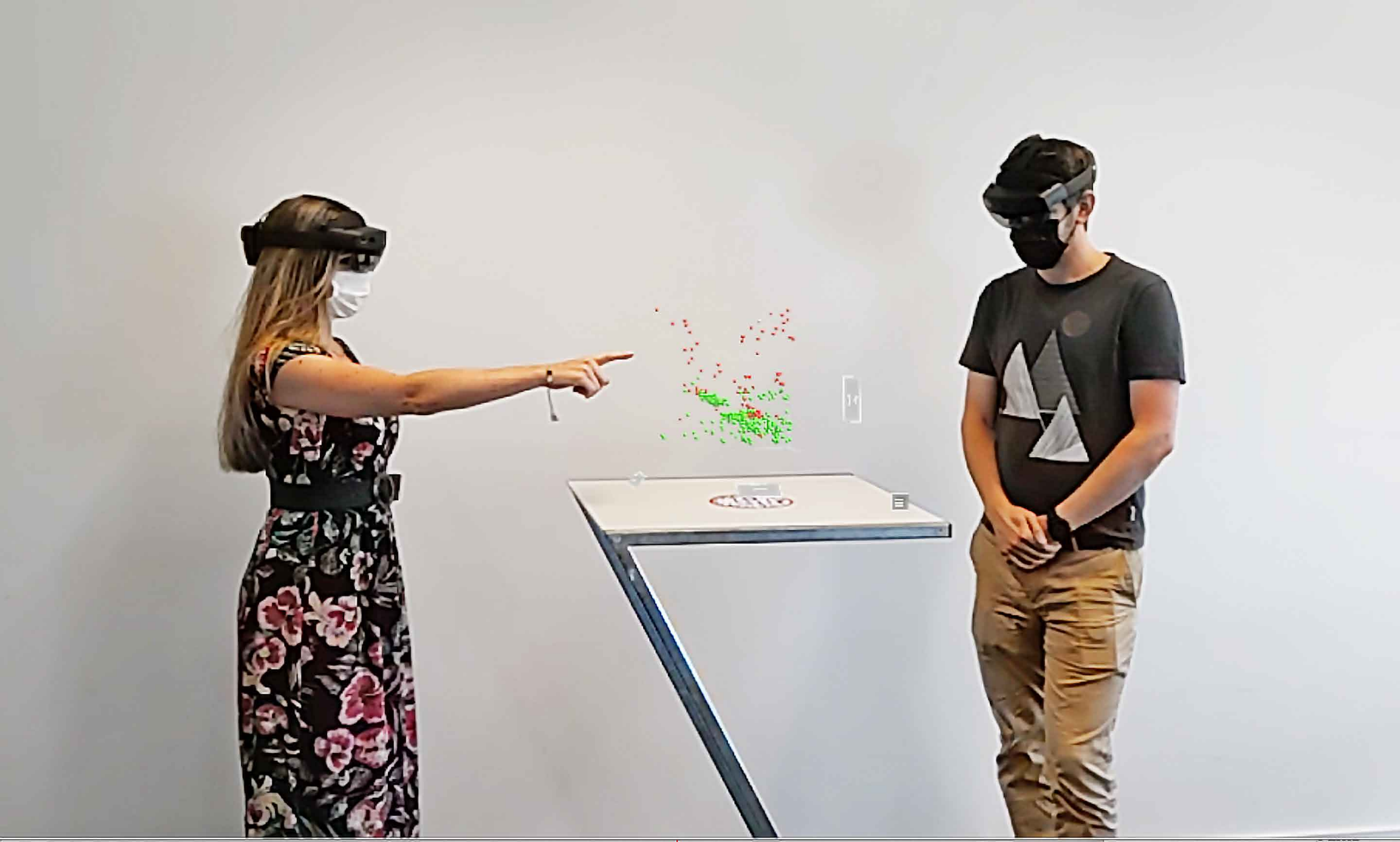} }}%
    {{ \includegraphics[width=0.503\linewidth]{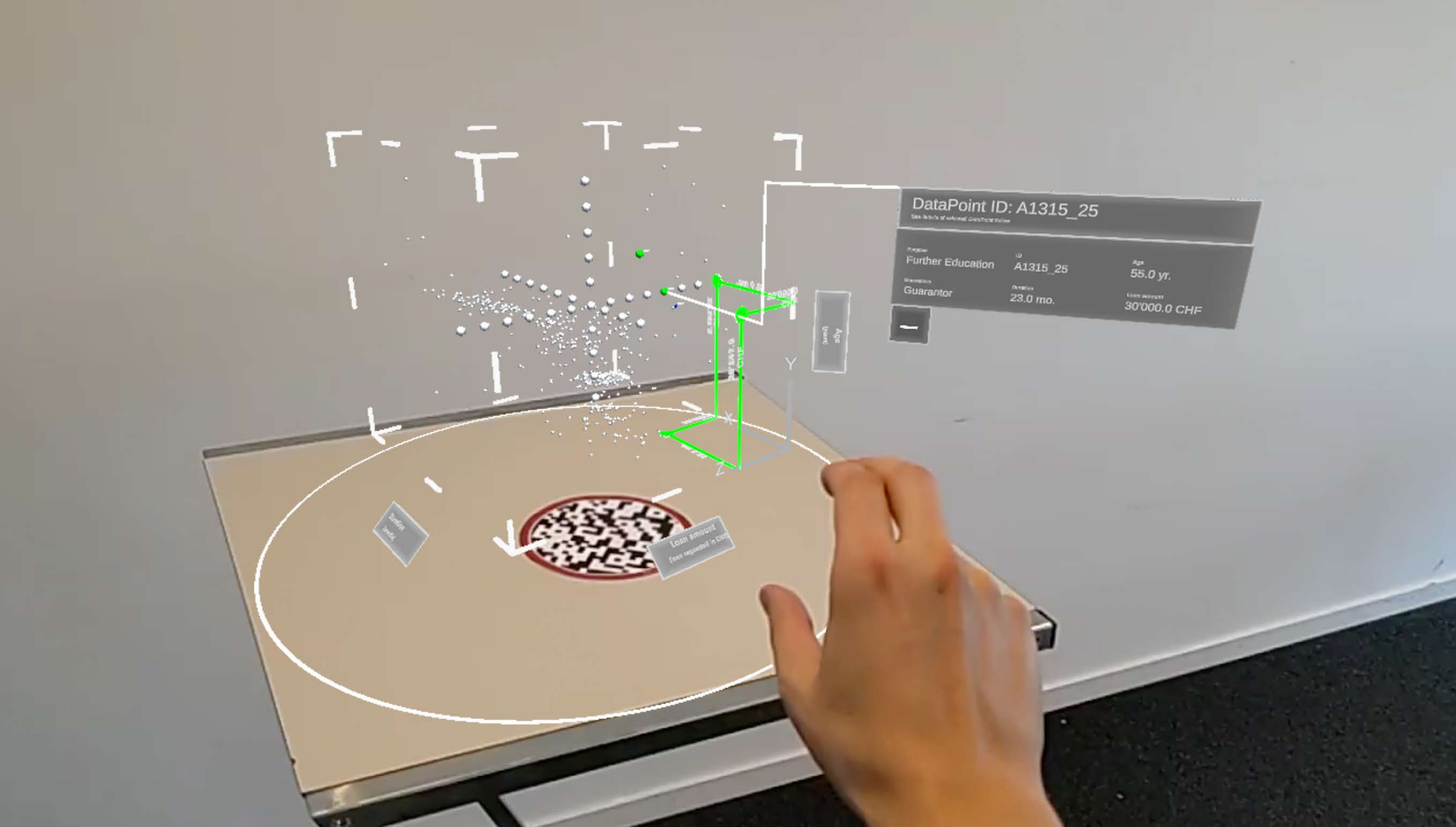} }}%
    \caption{
    AR application: the prediction explorer interface (left) allows pairs of users to interact with ML model predictions (colored data points) and to access additional information by air tapping on them. Users can also change the variables at each axis of the 3D scatterplot and access a menu with hand pointing. The visualization is synchronized between users. Counterfactual explanations (right) in the prediction explorer interface allow users to identify which changes a data point (represented by the three orthogonal lines consisting of bigger white spheres) should undergo  to change its current ML model prediction. The user can select different counterfactual data points. The color highlights the corresponding ML model outcome (green means that the credit request is ``accepted'').}%
    \label{fig:system_interfaces_prediction}%
\end{figure}

(2)~The prediction explorer (Figure~\ref{fig:system_interfaces_prediction}) visualizes an interactive 3D scatterplot of ML model predictions. Each sphere represents a data point that is colored according to the ML model prediction (two colors in the case of a binary classification model). The prediction explorer further comprises interactive labels for all axes, pop-up windows with metadata for all data points, and a menu comprising information in different tabs on the ML model performance measures. Furthermore, it contains the list of the current user tasks (see Section~\ref{sec:web_app}) and additional functionalities such as submitting results or moving the data cloud in the physical space. The main interactions in the prediction explorer are: selecting a data point (by air tapping), changing the dimension of a selected axis (by either touching the interactive labels or air tapping them), and interacting with the menu (by touching the buttons in the menu or air tapping them). The menu is designed to float next to the data cloud for direct access. Based on the selected model, the prediction explorer can also visualize 3D counterfactual explanations. This should help users to better understand how the ML model arrives at predictions. For example, in credit lending, one can answer questions, such as ``\emph{What is the minimum age for customer X that would have allowed him to get the requested loan, other things equal?}''. To generate counterfactual explanations, the prediction explorer allows users to move selected data points (by air tapping a new location along the selected dimension) and then it shows the prediction for the new values of the data point (see Figure~\ref{fig:system_interfaces_prediction}, right). 

A video demonstrating the different interactions and visualizations in the AR application can be found in the supplemental materials.

\subsubsection{Web application} 
\label{sec:web_app}

The web application simulates an email client in a browser that users access at desktop computers. Users can interact with the email client by providing input using a mouse and keyboard.

\subsubsection{Backend server} 

The backend server controls and updates the content visualized in the AR application (e.g., the dataset and ML models) and in the web application (e.g., emails). It controls, in particular, that all interfaces are synchronized.

\subsection{Room setup}

\begin{figure}[h]
    \centering
    \includegraphics[width=0.5\linewidth]{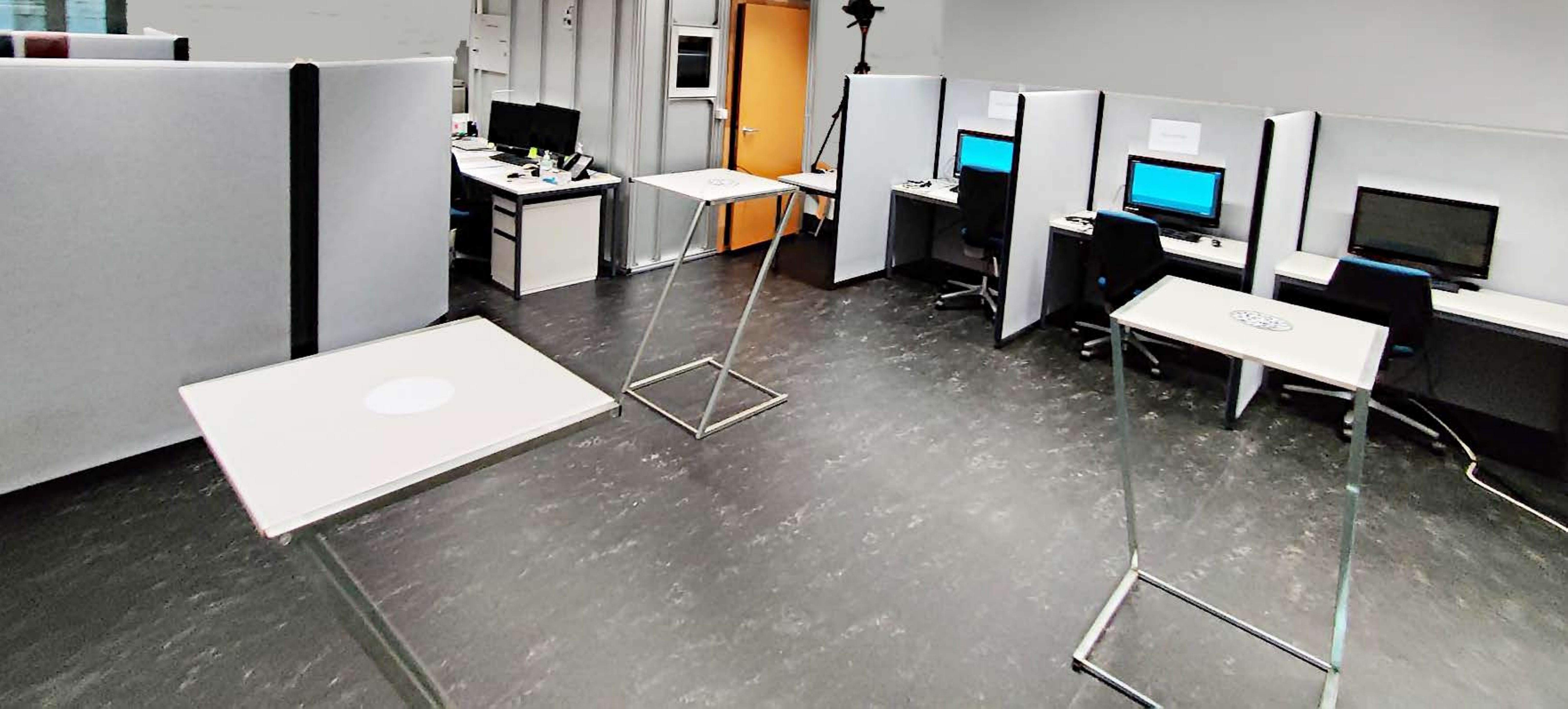}
    \includegraphics[width=0.332\linewidth]{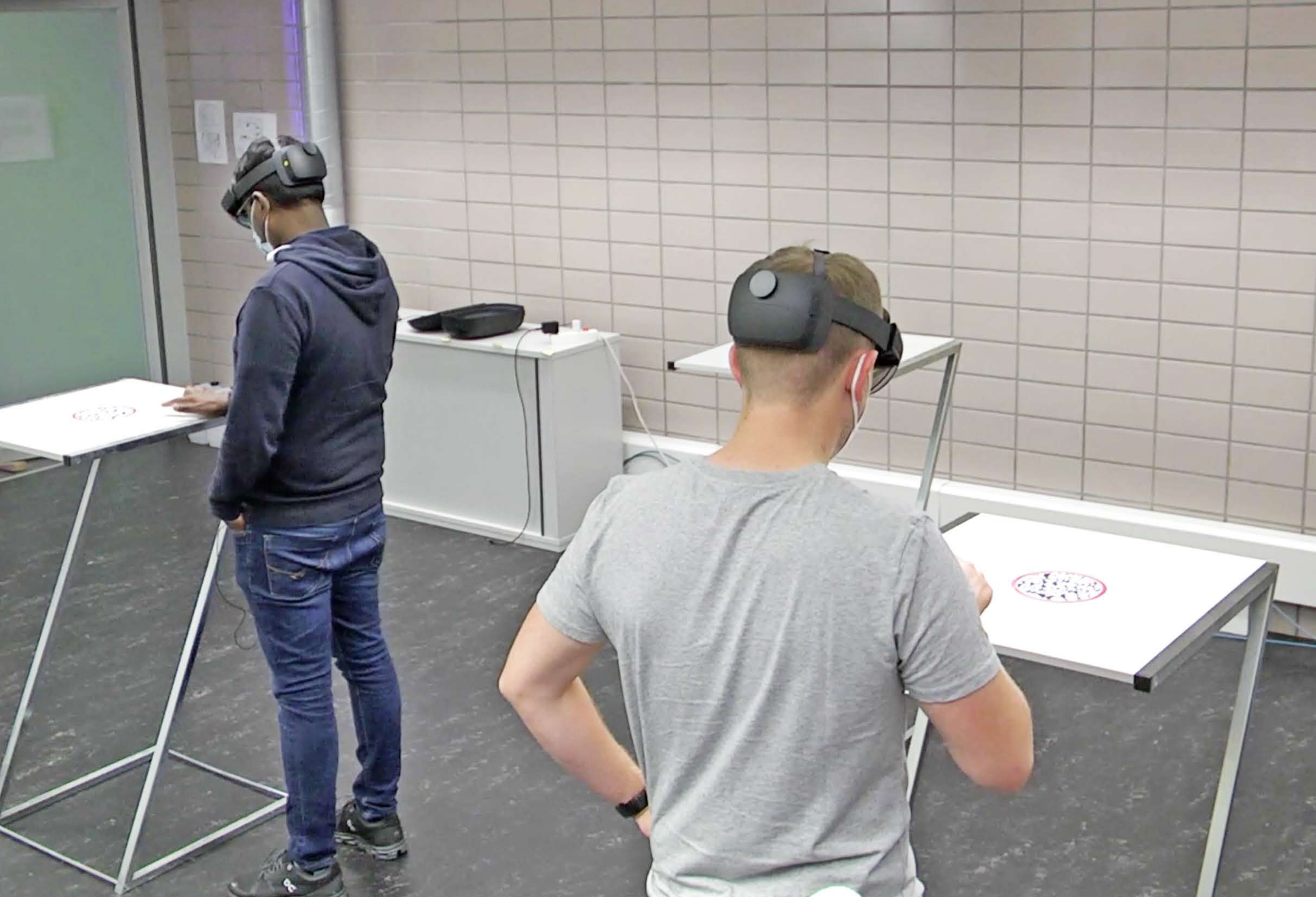}
    \caption{Experimental room setup with personal and shared workspaces (left). System users at their respective personal workspaces (right). The QR codes are visible on the tables.}
    \label{fig:system_setup}
\end{figure}

During our user study (see Section~\ref{sec:user_study}), we embedded the system in a physical space that is divided into three different workspaces  (Figure~\ref{fig:system_setup}): two personal workspaces (one per user) and a shared workspace. The personal workspaces both consist of a desk with a computer and a stand-up table with a height of 114 cm. The computer shows the web application running in a browser that is controlled by the backend server. On the stand-up table is a QR code that allows the AR application running on the HMDs to recognize the location (i.e., personal vs. shared workspace) and triggers the visualization right above the table surface. The system is designed to allow users to see and interact with the visualization at their own personal workspace. The height, size, and position of the tables allow users to walk around them and observe the visualization from all sides. The shared workspace also consists of a stand-up table with a QR code triggering the visualization above the table's surface. The visualizations shown at the shared workspace are synchronized between users: the system is designed to let users interact with the same visualization at the same time.

\section{User study}
\label{sec:user_study}

We tested our CIA system in a user study, where we examined the taskwork and teamwork dimensions of collaboration. We measured taskwork using performance assessments. We adapted the so-called pair analytics methodology \cite{AriasHernandez2011PairAnalytics} to our study setting. This allowed us to examine the emergence of teamwork among users with different professional backgrounds and levels of ML modeling domain knowledge while they solved industry-relevant ML tasks. Our user study was conducted in compliance with local COVID-19 guidelines and was approved by our Institutional Review Board.

\subsection{Participants}

We recruited 18 participants (5 female) of age 24 to 41 years for the study. The number of participants is in line with prior observational studies on IA applications  \cite{Wang2020TowardsAU, Batch2020ThereIsNoSpoon, Reski-et-al_2020, Spinner2020explAInerAV}. To participate in the study, participants needed to fill out a screening questionnaire, requiring them to be 18 years or older, fluent in English, have at least basic (self-assessed) ML knowledge, and not suffer from any binocular vision disorder or wear a personal medical device, such as a pacemaker, to mitigate the risks associated with wearing HMDs. 

In the study, we asked participants to adopt the role of either a risk officer or a ML engineer. Therefore, we recruited participants with different backgrounds. Specifically, participants in our user study were (1) industry professionals, and (2) university students. For the sake of brevity, we refer to the industry professionals as ``IND'', and to the university students as ``STU'', respectively. The industry professionals (9 participants) were recruited from a Master for Advanced Studies (MAS) program organized at the local university. The industry professionals were all enrolled in a block course of the MAS program covering the topic of ML for industry executives. The block course was conceived for professionals seeking to improve their education on industry-relevant ML applications. Out of 9 IND in the study, 5 indicated to be in management-related roles, 3 in business or product engineering roles, and 1 indicated to be a technical architect. The university students (9 participants) were recruited via a mailing list from the pool of master's and PhD students of the local university, where they all completed Master of Science (MSC)-level courses on marketing or business analytics within the previous two years. These courses focus on ML modeling in business contexts. Participation in these courses ensured a broad knowledge of machine learning that we deemed necessary to take the role of a ML engineer in this study. STU were renumerated with a giftcard of an equivalent of around \$20 . IND study participation was linked to the grading of the block course, as they were asked to write a report on the user study, highlighting possible applications of CIA in industry. 

\subsection{Study design and procedure}
\label{sec:Study_Design_Proc}

\begin{figure}[h]
    \centering
    \includegraphics[width=\linewidth]{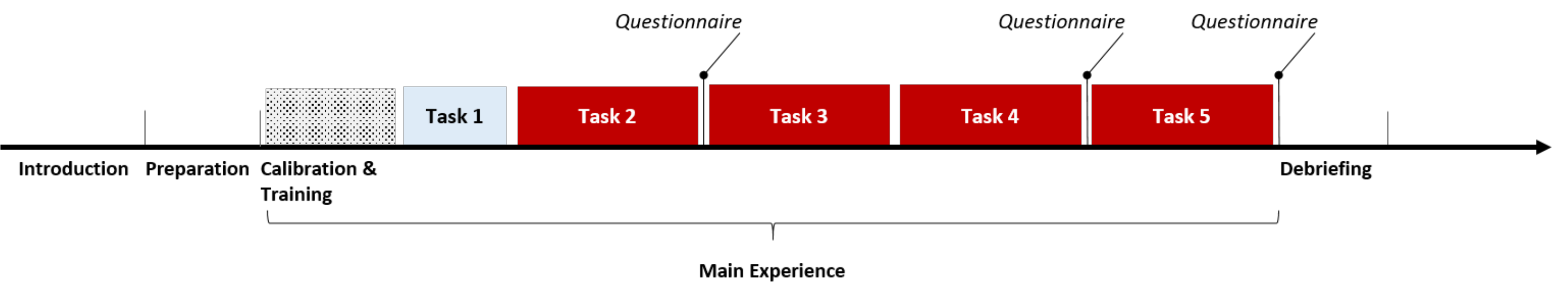}
    \caption{Procedure of the user study. We highlight the sequence of study blocks and the collection of the users' feedback at the end of tasks 2, 4, and 5. With the exception of task 1, all tasks in the experience block required collaboration among the users.}
    \label{fig:study_procedure}
\end{figure}

Each study session consisted of four main blocks: \textbf{introduction}, \textbf{preparation}, \textbf{experience}, and \textbf{debriefing}. The duration of the experience block was 30 to 60 minutes, and the entire user study lasted about 90 to 120 minutes in total. Two participants, i.e., one IND and one STU, were matched in each session. All study participants attended only one session. We summarize the study procedure in Figure~\ref{fig:study_procedure}.

\vspace{0.1cm}
\noindent\textbf{Introduction.} The introduction took place in a meeting room. Both participants were asked to read an information sheet, as well as fill out a consent and COVID-19 contact tracing form. One of the experimenters familiarized the participants with the purpose of the study, the procedure, the setup of the room for the experience, and the roleplay storyline. Specifically, participants were introduced to a realistic banking scenario on algorithmic credit lending, i.e., where ML is used to assess the creditworthiness of potential customers. In the scenario, each participant was assigned to one of two roles based on their knowledge background: the IND was assigned the role of a risk officer, while the STU was assigned the role of a ML engineer. In the storyline of the banking scenario, participants received ML tasks from fictitious line managers where the goal was to examine and improve the ML model for credit lending at the bank. 

We used an adapted version of the German credit lending dataset\footnote{\url{https://archive.ics.uci.edu/ml/datasets/statlog+(german+credit+data)}} in the AR application. The dataset consists of tabular data and comprises features that are relevant for credit lending and risk management, such as credit and savings amount, the purpose of the loan, and the age of the customers. We intentionally chose credit lending as it is commonly used in the ML literature, especially for industry-relevant applications of ML, as well as to discuss the generation of counterfactual explanations \cite{Wachter2017CounterfactualEW,lipton2018mythos}.

\vspace{0.1cm}
\noindent\textbf{Preparation.} In the preparation block, the participants were brought to the experiment room and introduced to its layout. They were asked to watch a short video giving them a brief refresh of selected ML concepts. Afterward, the participants were asked to complete a short questionnaire, assessing their demographics, their knowledge of ML and counterfactual explanations, and their past experiences with immersive technologies (see Section~\ref{sec:measures}). 

\vspace{0.1cm}
\noindent\textbf{Experience.} At first, the participants performed the eye-tracking calibration of the HoloLens 2 devices and were familiarized with air tapping and hand pointing through a short training. Then, the experience with the system started. Participants were asked to perform five different ML tasks sequentially. Each task started with the participants receiving an email from the respective line managers on the web application running on the computers at their personal workspace, giving them individual instructions (based on their role) on how to proceed with the experience. Instructions were designed to provide participants with a task structure that did not limit them in their ability to collaboratively explore various choices and aspects of the system, even though they never used the system before. The proposed tasks and their order match common practice in risk management for financial institutions, such as a bank, where well-established ML models are periodically validated and managers collaborate with data scientists to introduce new models. We briefly summarize the tasks below.

\noindent\emph{Task 1.} Participants explored visualizations at their personal workspaces independently, without talking to each other. This was done to familiarize themselves with their role and the overall storyline. Specifically, they were asked to use the prediction explorer in the AR application to explore the ML model for credit lending  currently in use at the bank. 

\noindent\emph{Task 2.} Participants received a prompt in the AR application asking them to explore the visualization together at the shared workspace. The goal of the collaborative task was to let the participants better understand and identify potential problems of the ML model for credit lending. At the end of task~2, the participants filled out a short questionnaire using the web application at each personal workspace to assess their understanding of the data and the ML model.  

\noindent\emph{Task 3.} Participants were asked to follow a set of complementary requirements provided by their line managers and train an interpretable ML model from the family of generalized linear models (GLM) \cite{nelder1972generalized} together at the shared workspace.\footnote{The risk officer (IND) was asked by the line manager to consider GLMs, as these models are currently used in financial institutions for quantitative risk management. The line manager of the ML engineer (STU) contributes by introducing additional requirements for training the GLMs.} To do so, they had to use the model designer of the AR application and select the correct ML problem type, classifier family, and variables to be used, among others.   

\noindent\emph{Task 4.} Participants were requested by their line managers to validate the GLM they trained in task~3 at the shared workspace. To do so, they were asked to analyze how the performance measures, such as accuracy, sensitivity, and specificity, vary by changing the classification threshold. Finally, participants were asked to individually fill out a questionnaire at their personal workspaces. 

\noindent\emph{Task 5.} Participants were now asked to train a ``black-box model'', i.e., a neural network,\footnote{The use of a ``black-box model'' is suggested by the line manager of the data science officer (STU). This reflects what commonly happens in financial institutions such as banks, where interpretable models are increasingly replaced with black-box models, due to their better performance. The line manager of the risk officer (IND) introduced additional requirements for model training.} following a set of complementary requirements provided by their line managers. After training, they were asked to use counterfactual explanations in the AR application to answer a set of business-relevant questions (e.g., ``\emph{What is the maximum age at which customer B1821 would get a loan, other things equal?}''). Finally, they individually filled out a questionnaire at their personal workspaces. 

At the end of the experience, participants took off the HMDs and filled out a final set of questionnaires, assessing the system usability and user experience.

\vspace{0.1cm}
\noindent\textbf{Debriefing.} At the end of the study, we conducted a short debriefing with each participant, where we asked them if they wanted to provide any additional qualitative feedback or remarks about the experience, or if they had any additional questions. 

\subsection{Measures}
\label{sec:measures}

\noindent{\textbf{Collaboration: Teamwork assessment}}.
We evaluated the teamwork component of collaboration qualitatively and quantitatively by video- and audio-recording participants' interactions throughout the experiment with a camera located in the corner of the experimental room and microphone clips for each of the participants. We adapted the pair analytics methodology \cite{AriasHernandez2011PairAnalytics} from collaborative visual analytics to our user study. We subsequently coded the recordings to identify user-focused trends and issues of relevance for future work in CIA. The recordings of 9 sessions were independently coded by two experimenters, using the linguistic annotation tool ELAN.\footnote{https://archive.mpi.nl/tla/elan} The two experimenters compared the results of the coding of each session and verbally resolved disagreements.

The pair analytics methodology consists of an experimental protocol and a suggested coding scheme introduced by \citeauthor{AriasHernandez2011PairAnalytics} \cite{AriasHernandez2011PairAnalytics}. Importantly, we can later use the coding for quantitative analyses. Pair analytics relates to protocol analysis (or the ``think-aloud'' method) that is commonly used in human-computer interaction research, although in a collaborative setting. However, it offers a less invasive way of capturing users’ thought processes \cite{AriasHernandez2011PairAnalytics}. In fact, the think-aloud method may influence or bias participants' thought processes \cite{Wang2020TowardsAU, Kuusela2000ACO}, or may not capture their thought processes accurately \cite{Charters2003TheUO}. As our experimental protocol follows the pair analytics methodology, participants in our study interacted with no interruption. 

Central to the pair analytics method is the use of two roles, i.e., a subject matter expert (SME) and a visual analytics expert (VAE) \cite{AriasHernandez2011PairAnalytics}. In our study, the SME is a risk officer, played by the IND, and the VAE is a ML engineer, played by the STU, as described in Section~\ref{sec:Study_Design_Proc}. Our adaption of the methodology differs from the original method in that the IND and STU---while closely matching the original description of the roles---are nevertheless proxy users.

Following \citeauthor{AriasHernandez2011PairAnalytics} \cite{AriasHernandez2011PairAnalytics}, we employed a coding scheme for the video and audio data based on the theory on joint actions that allowed us to capture the  participants' joint actions and the processes used to coordinate them \cite{Clark1991GroundingIC, AriasHernandez2011PairAnalytics}. We considered three main code groups: (1) coordination of attention (\texttt{COA}), (2) pauses in joint actions (\texttt{PJA}), and (3) navigation of joint actions (\texttt{NJA}). We subsequently used the codes to analyze patterns of successful collaboration in immersive analytics when performing ML tasks. Following \citeauthor{Billinghurst2018CollaborativeIA} \cite{Billinghurst2018CollaborativeIA}, we grouped the patterns into six main dimensions of collaborations, called topics, and mapped the codes from the video analysis to the topics they best characterized. The resulting topics are as follows. 

\vspace{0.1cm}
\noindent\emph{(1) Group (pair) dynamics.} Group dynamics can take on various forms and depends on various factors, such as context, diversity, or group size \cite{Heer2008DesignCF}. In the context of CIA, one of the influencing factors may be that facial expressions from collaborators are hidden when wearing HMDs  \cite{Billinghurst2018CollaborativeIA}. Here, questions of interest are how pairs establish their roles when manipulating the visualization, and how wearing HMDs may affect collaboration. We used the code group \texttt{NJA} to explore these questions.

\vspace{0.1cm}
\noindent\emph{(2) Division and allocation of work.} Dividing and allocating tasks is a central component of successful collaboration \cite{Heer2008DesignCF}. Although the modularity of participants' work segmentation in our study was largely dictated by the tasks in the role assignment, we focused on several topics regarding sense-making and problem-solving. These include how participants handled the navigation of tasks, whether the tasks were divided equally, and how task division affected participants' individual and joint performance. We used the code group \texttt{NJA} to analyze these topics. 

\vspace{0.1cm}
\noindent\emph{(3) Common ground and awareness.} Common ground denotes a shared understanding between conversational partners, enabling successful communication and collaboration \cite{Clark1991GroundingIC, Roschelle1995TheCO}. Awareness ``is an understanding of the activities of others, which provides a context for your own activity'' \cite{Dourish1992}. Both common ground and awareness are affected by how an augmented space is spatially defined and used by participants, by the level of artificiality and immersion, and by cognitive overload or distractions \cite{Sereno2020CollaborativeWI}. Furthermore, seeing the same visual environment is key for successful collaboration \cite{Heer2008DesignCF}. Thus, we were interested in understanding when common ground was easier or more difficult to establish, and how shared situational awareness in terms of activities and group members \cite{Hong2018CoordinatingAP, Scott2015LocalRC, Paul2010UnderstandingTS} was affected by our system. All code groups were used to analyze this topic.\\ 

\vspace{0.1cm}
\noindent\emph{(4) References and deixis.} Spatial referencing and deixis are central components for establishing common ground among participants \cite{Billinghurst2018CollaborativeIA, Heer2008DesignCF}. Examples of spatial referencing and deixis are pointing, verbal markers, or moving objects to direct attention \cite{Clark1991GroundingIC}. We used the code group \texttt{COA} to explore how participants made use of references and deixis during the experience with our CIA system.

\vspace{0.1cm}
\noindent\emph{(5) Incentive and engagement.} Collaborators' incentives to engage in a task can vary greatly depending on the context in which collaboration takes place \cite{Heer2008DesignCF}. As our study was conducted in a laboratory environment with fixed roles, we focused on participants' hedonic incentives, i.e., ``well-being [...] experienced intrinsically in the work'' \cite{Heer2008DesignCF}. Engagement is ``the emotional, cognitive and behavioral connection that exists, at any point in time and possibly over time, between a user and a resource'' \cite{Attfield2011TowardsAS} and can potentially lead to a state of ``flow'' \cite{Sereno2020CollaborativeWI}. We analyzed pauses in joint actions (\texttt{PJA}) to observe whether participants' engagement in the ML tasks increased or decreased over time, and whether this was in relation to cognitive workload. 

\vspace{0.1cm}
\noindent\emph{(6) Consensus and decision-making.} In real-world ML settings, domain experts and ML engineers may need to reach consensus and make joint decisions, such as whether or not to replace existing ML models with others, or whether they agree on patterns in data. We designed the ML tasks to place pairs of users in a realistic ML setting, allowing us to observe whether and how they reached consensus and then made decisions to progress with the assigned ML tasks. We used the code group \texttt{NJA} to learn about relevant patterns.

Additionally, we measured \emph{interaction times} with the visualization using the code groups \texttt{NJA} and \texttt{PJA}, in order to quantitatively assess how much time was spent actively and passively interacting with the visualization at the shared workspace. We define \emph{active visualization interaction} as time spent manipulating the visualization, and \emph{passive visualization interaction} as time spent discussing or observing the visualization. We measure the amount of time IND and STU (1) individually manipulate the visualization during the collaboration; (2) jointly and actively manipulate the visualization; and (3) jointly and passively observe or discuss the visualization. 

\vspace{0.1cm}
\noindent{\textbf{Collaboration: Taskwork assessment.}} We assessed the taskwork component of collaboration \cite{Ens2021GrandCI} by quantitatively evaluating participants' task performance. In our user study, this is given by the number of correct answers to task-related questions at the end of tasks 2, 4, and 5. At the end of task 2, both participants filled out a joint questionnaire. At the end of tasks 4 and 5, they completed separate questionnaires instead. Furthermore, we evaluated participants' confidence in the ML models and counterfactual explanations, and the willingness to replace an existing ML model. For this, we used a set of questions at the end of tasks 2, 4, and 5. Here, participants rated (1) their willingness to replace a biased ML model (task~2), (2) their confidence in the performance of a ML model compared with an alternative one (task~4), and (3) their confidence that the counterfactual explanations can provide understandable explanations on how a customer with a denied loan application could get a loan (task~5). We refer to the supplemental materials for details.

\vspace{0.1cm}
\noindent\textbf{Usability and user experience.} We made the following quantitative assessments: we assessed usability and user experience using the system usability questionnaire \cite{brooke1996sus} and task load using the NASA TLX raw \cite{Hart2006NasaTaskLI}. We further asked user feedback questions that allowed participants to state what they liked, disliked, would have changed in the system, and their opinion on the applicability of such systems in actual business applications (see supplemental materials).

\section{Results}

\textbf{Collaboration: Teamwork assessment}.
The qualitative assessment of the audio and video recordings yielded the following results for the six topics of collaboration. 

\begin{figure}[h]
    \centering
    \includegraphics[width=\linewidth]{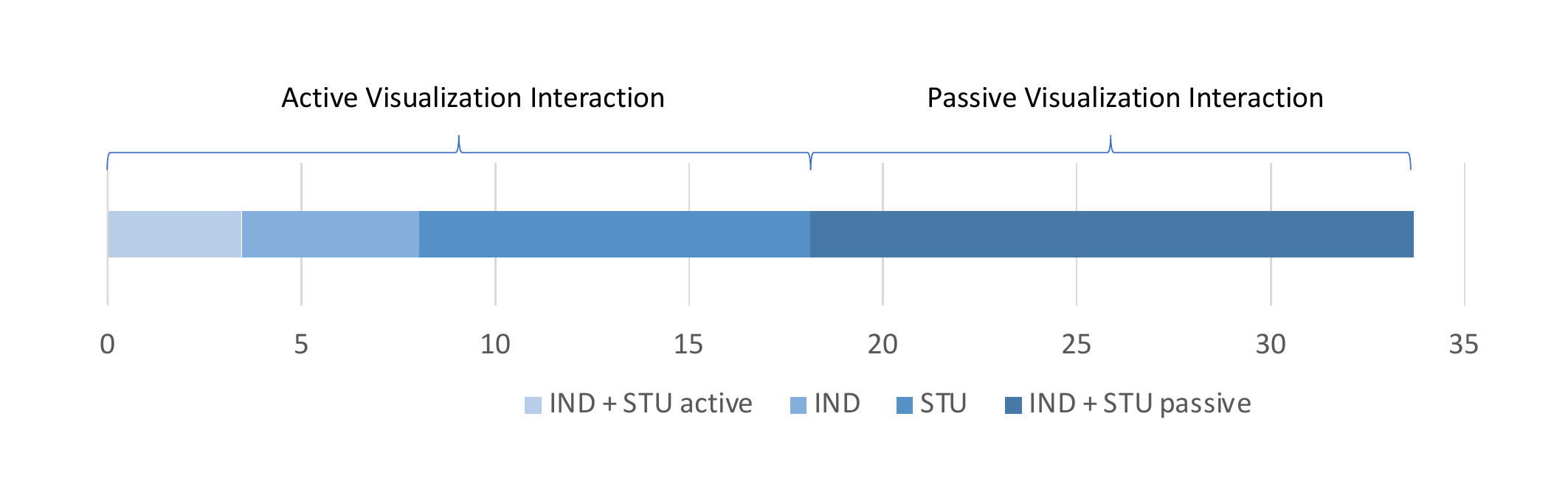}
    \caption{Time spent interacting with the visualization. "IND+STU active" signifies time spent actively and simultaneously manipulating the visualization, while "IND+STU passive" signifies time spent observing and discussing the visualization. Colors also show "IND" and "STU" separately, which signifies time spent individually leading the manipulation of the visualization.}
    \label{fig:time}
\end{figure}

\vspace{0.1cm}
\noindent\emph{(1) Group (pair) dynamics.} 
Although the levels of active and passive engagement varied across participants, in all sessions, one of the participants took on the role of managing the navigation of ML modeling tasks (usually the IND). This was typically indicated by a greater number of vertical\footnote{The terms ``vertical'' and ``horizontal'' transitions are used by \citeauthor{Bangerter2003NavigatingJP} to denote the entering and exiting, or continuation of a collaborative activity, respectively \cite{Bangerter2003NavigatingJP}.} transitions when navigating joint actions (e.g., ``\emph{Okay, let's move to the next task}''). The other participant took on the role of driving and executing the tasks and, therefore, moving along activities horizontally (e.g., ``\emph{Mhm}'', ``\emph{Sure, let's do that}'').\footnote{In pair programming, the terms ``navigator'' and ``driver'' are commonly used for these roles \cite{AriasHernandez2011PairAnalytics, Chong2007TheSD, Jung2012GroupHB}, which we will employ for the remainder of this work.} Exceptions were session S4, where the STU took on both driver and navigator roles, and session S3, where both the IND and STU switched between roles. Additionally, on average, STU spent more time actively manipulating the visualization (mean 10.0 minutes, standard deviation [SD] 4.6), while IND spent less time (mean 4.6, SD 4.2). This is shown in Figure~\ref{fig:time}.

We also noted a general turn-taking behavior among participants during their interactions with the visualization, with ``handovers'' similar to that of interfaces, such as a keyboard and mouse, between two users at a desktop computer. Despite the fact that participants could interact simultaneously with the visualization at the shared workspace, they only did so approximately 10 \% of the interaction time, or on average 3.5 minutes (SD 2.1), as seen in Figure~\ref{fig:time}. Overall, there were on average 7.9 (SD 3.0) handovers per session. On average, STU received control 2.8 times (SD 1.6) and IND 5.1 times (SD 2.3) per session. In several sessions (S1, S2, S3, S6, S7), handovers were characterized by asking for consent before interacting with the visualization. Consent was usually given by the person in the ``driver'' role. In session S3, for instance, the IND uttered: ``\emph{Hm, that's interesting ... uhm, can I ... check some?}'' and the STU answered: ``\emph{Yes sure, go ahead!}''.

\vspace{0.1cm}
\noindent\emph{(2) Division and allocation of work.} Overall, most pairs discussed the tasks openly and divided the work in a fluent manner. In several sessions (S1, S6, S7, S8, S9), the IND navigated actions by asking the STU for elaborations or by instructing the STU what to do, thereby monitoring and managing the collaborative tasks. The STU took charge of executing requests and manipulating the visualization, answering questions, or offering unsolicited elaborations. In line with the role division discussed in (1)~\emph{Group (pair) dynamics}, task division was implicitly agreed upon. The following dialogue extract exemplifies an STU in the driver role, elaborating on a data point, and directing IND's attention to it:
\begin{dialogue}
\small
\speak{STU} \emph{His} [of a customer] \emph{savings are actually not that much?}
[IND moves around the shared workspace table towards STU. Both pause to look at the visualization.]
\speak{STU} \emph{Yeah, so his savings are rather low.}
\speak{IND} \emph{Which ... where is this?}
\speak{STU} \emph{He is the ... flashing point ...} [points at the visualization]
\speak{IND} \emph{Ah yeah.}
\speak{STU} \emph{Is it also flashing for you?}
\speak{IND} \emph{Yeah, yeah.}
\end{dialogue}
In a few sessions with less clear roles (S3, S4), the participants decided more explicitly on who should perform a task. In sessions S2 and S9, we noticed that, when participants failed to mention the details of their tasks, the lack of information hindered collaboration and led to confusion, as signified by an increased number of questions and elaborations.

\begin{figure}[h!]
    \centering
    {{ \includegraphics[width=0.312\linewidth]{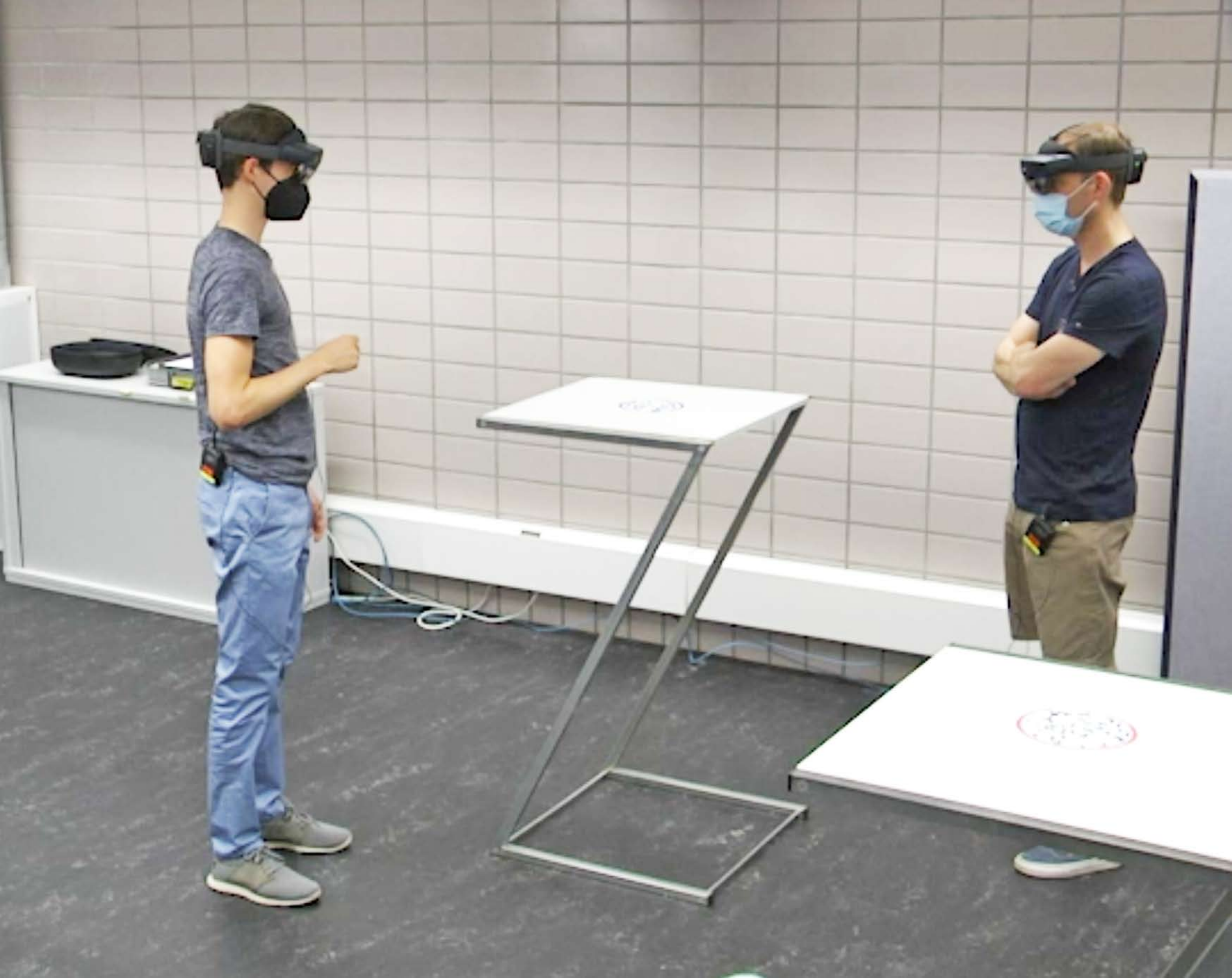} }}%
     {{ \includegraphics[width=0.3\linewidth]{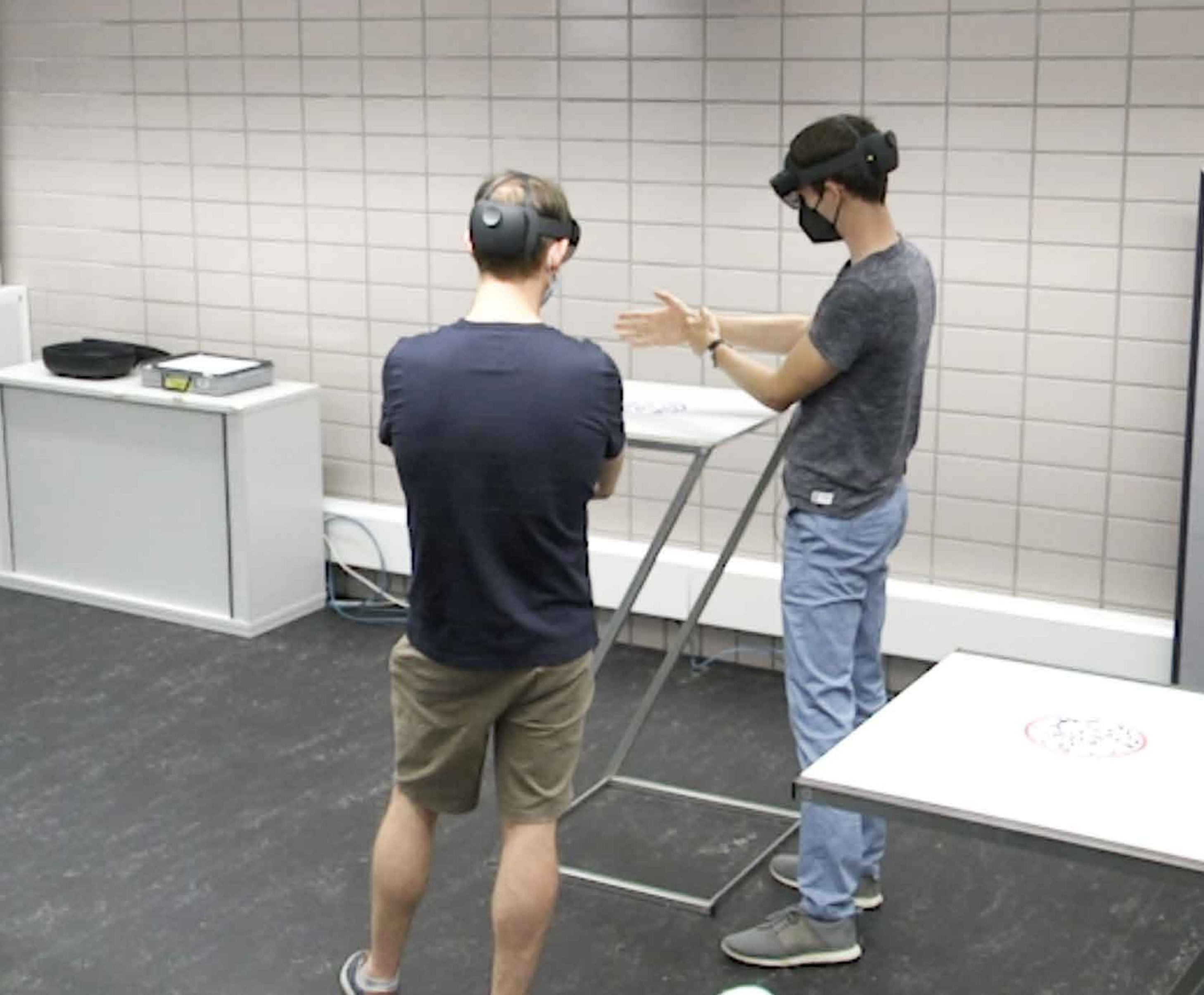} }}%
    \qquad
    \caption{Participants establishing collaboration at the shared workspace (left). STU trying to visualize a plane using both hands to explain a pattern in the data to IND (right). }%
    \label{fig:system_colab1}%
\end{figure}

\vspace{0.1cm}
\noindent\emph{(3) Common ground and awareness.} Overall, most pairs offered unsolicited elaborations on the data visualizations, ML tasks, or their thought processes to help establishing common ground. At the beginning of each collaborative task, the  participants usually stood on almost opposite sides of the holographic visualization, facing each other and creating a personal space around themselves (S1, S2, S4, S5, S6, S7, S8, S9), as shown in Figure~\ref{fig:system_colab1} (left).  

\begin{figure}[h!]%
    \centering
    {{ \includegraphics[width=0.3122\linewidth]{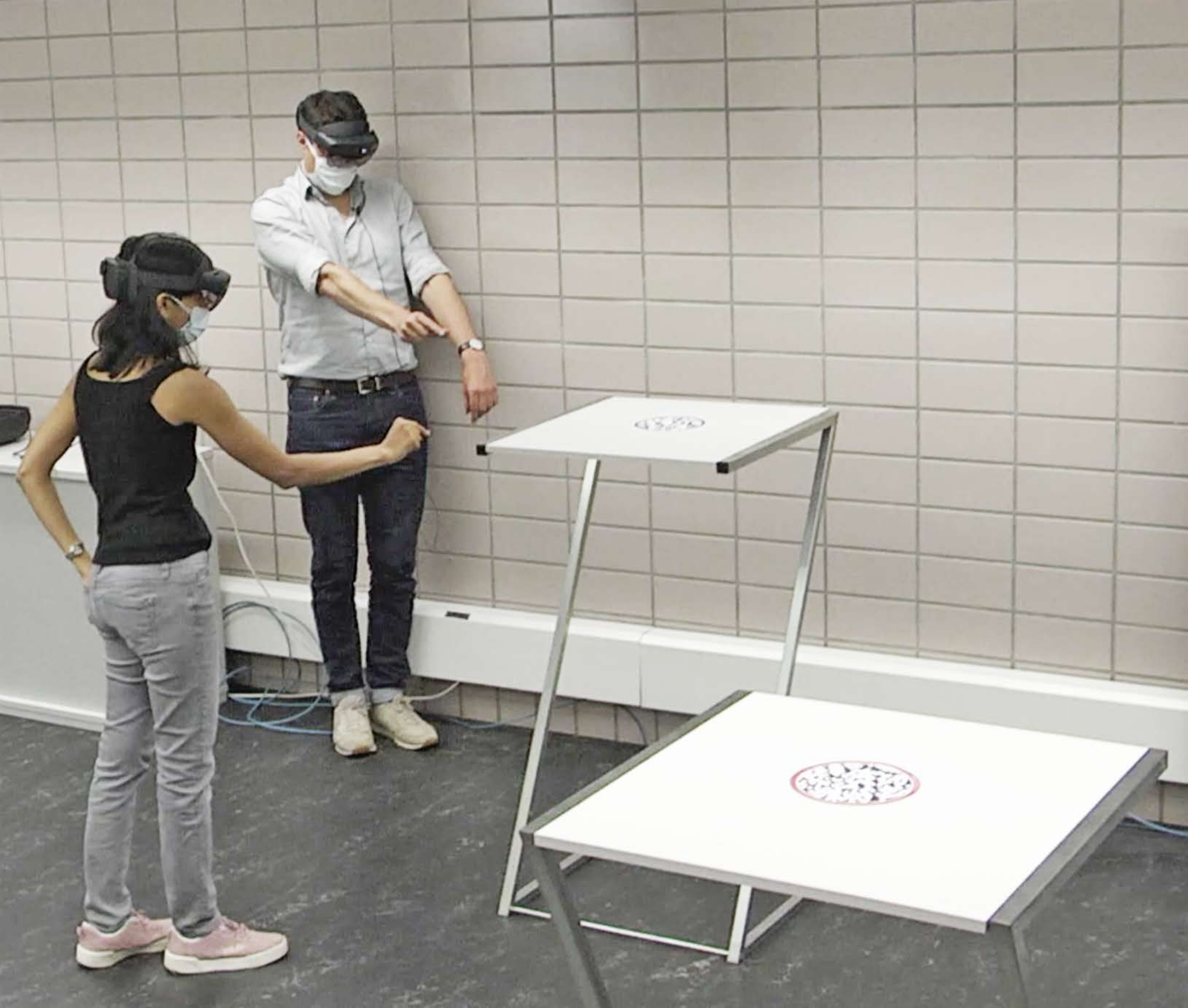} }}%
    {{ \includegraphics[width=0.312\linewidth]{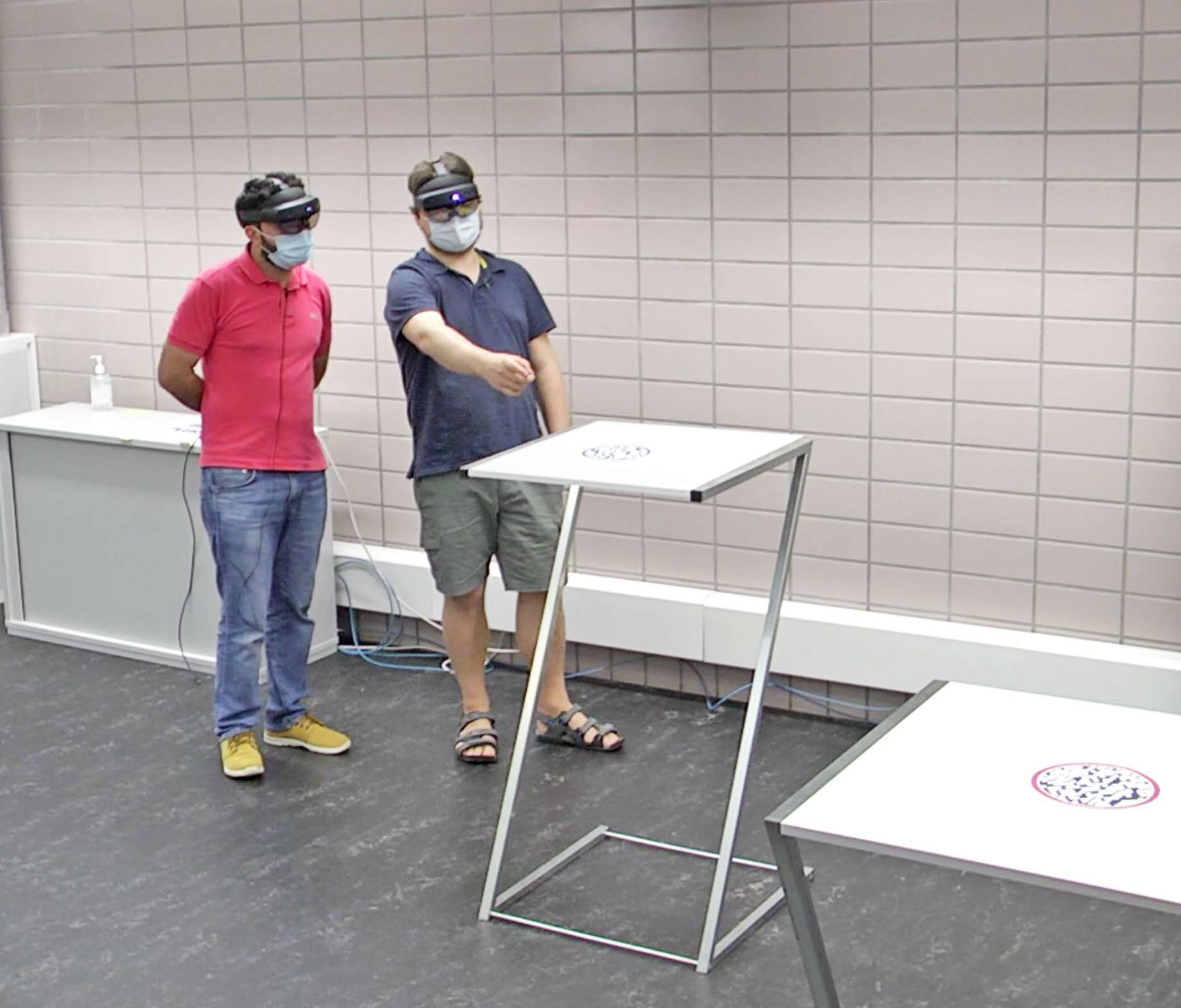} }}%
    \qquad
    \caption{Participants updating their spatial awareness at the shared workspace: STU has reached the wall of the experiment room (left). STU and IND discussing the visualization by pointing at data points: IND moves next to STU and gains a closer view of STU's reference point to enable taskwork (right).}%
    \label{fig:system_colab2}%
\end{figure}

Participants sometimes moved closer to each other during the interactions at the shared workspace, as shown in Figure~\ref{fig:system_colab2}. For instance, once one of the participants directed attention to a point in the visualization, e.g., by uttering ``\emph{I mean the point at the top}'' or by a pointing gesture, the other participant moved closer to the first to get a closer view of the speaker's reference point, as in Figure~\ref{fig:system_colab2} (right). In most of the sessions, the ``navigator'' mostly followed the ``driver''. 

In all sessions, participants checked regularly with each other if their visualizations at the shared workspace matched at a given time, even though they had been told by the experimenters in the introduction block that the visualization at the shared workspace was synchronized and shared among participants. To do so, they questioned, confirmed, or negotiated joint attention. They either asked each other directly, with questions, such as ``\emph{Do you also have age on the z-axis?}'', or by verbally sharing what they saw in their HMDs, which occurred on average 9.1 (SD 4.3) times per session. Re-evaluating their common ground with respect to the visualizations often lead to interruptions in the flow of solving the task at hand. 

In a few sessions (S6, S7, S9) much of the time spent actively and simultaneously manipulating the visualization (see Figure~\ref{fig:time}) was spent in silence. This was particularly prominent in sessions S6 and S7, where 90.1 \% and 87.4 \% of this time was spent silently. Furthermore, in some sessions (S3, S4, S6), the lack of common ground with respect to the task was characterized by longer pauses in joint actions among participants and self-talk. These silent single-user interactions usually ended because of a new elaboration or question by one of the participants. Finally, participants often shared also the desktop computers at the personal workspace, acting as they were in a realistic ML setting to address taskwork by sharing their understanding. An example is given in the following dialogue:
\begin{dialogue}
\small
\speak{IND} \emph{So now ...}
\speak{IND} [self-talk] \emph{Now open ... no it says ...}
\speak{STU} \emph{What's the task?} [...] \emph{How to optimize the performance.}
[of the ML model] [Participants pause, both silently interact with the visualization at the shared workspace]
\speak{STU} \emph{Ok so let me double-check what it says over here.} [STU walks to the computer at the personal workspace. IND follows STU]
\speak{STU} \emph{Ok, so} [the e-mail of the STU line manager says] \emph{we should discuss how to optimize the performance.}
\speak{IND} \emph{Ok, you can have a look at my screen, if you want.}
\speak{STU} \emph{Thank you.} [walks to IND's computer at the personal workspace and reads the email]
\end{dialogue}

\normalsize

\noindent\emph{(4) References and deixis.} Participants employed a mixture of pointing and verbal markers (e.g., ``\emph{there}'', ``\emph{here}'') to channel the listeners' attention and to ensure that the visualization was synchronized among both participants. This is shown in Figures \ref{fig:system_colab1} (right) and \ref{fig:system_colab2} (right). Overall, we observed a similar amount of pointing and verbal markers, as well as a similar use among participants (see Table~\ref{tab:pointingTable}). In a few sessions (S1, S2, S6), the  participants visualized their elaborations by gesturing directly in the visualization, rather than pointing at a given data point. For example, the STU in session S1 elaborated on a general pattern in data, while trying to visualize a plane using both hands (seen in Figure~\ref{fig:system_colab1}). In most sessions, pointing was primarily employed in the collaborative exploration (S1, S2, S4, S6, S9) and decreased in the later stages of the experience.

\vspace{0.1cm}
\noindent\emph{(5) Incentives and engagement.} Overall, we found that all pairs of participants were engaged and immersed in the visualizations. We noted a slight overall decrease in incentive and engagement toward the end of the sessions (after approximately 30 minutes in the experience). We confirmed this in two ways: (1) We found a decrease of content-related discussions and an increase in discussions on the functionality of the system (e.g., how to interact with the selected components of the visualization). (2) We found an increase in the extent of \texttt{PJA}. In addition, there were signs of fatigue: in the last task of session S7, for instance, participant P2 said ``\emph{I cannot click anymore}'' to indicate physical fatigue, also referred to as gorilla arm \cite{HincapiRamos2014ConsumedEA}. 

\begin{table}[h]
\caption{Deixis (DX) throughout the collaboration. Stated: mean (standard deviation) of pointing (PT) and verbal markers (VB) used.}
\begin{tabular}{l|l|l|l|l|l|l}
 \toprule
 \emph{Total} &
\multicolumn{2}{c|}{\emph{DX Both}} &  \multicolumn{2}{c}{\emph{DX STU}} &  \multicolumn{2}{|c}{\emph{DX IND}} \\
\hline
& VB & PT &VB & PT &VB  & PT\\ 
\hline
    
\multicolumn{1}{r|}{50.7 (14.8)} &
  \multicolumn{1}{r|}{27.7 (10.3)} &
  \multicolumn{1}{r|}{23.0 (6.4)} &
  \multicolumn{1}{r|}{16.4 (7.2)} &
  \multicolumn{1}{r|}{12.9 (7.2)} &
  \multicolumn{1}{r|}{11.2 (7.2)} &
\multicolumn{1}{r}{10.1 (4.0)} 

  \\  \bottomrule
\end{tabular}
\label{tab:pointingTable}
\end{table}

\vspace{0.1cm}
\noindent\emph{(6) Consensus and decision-making.} We found that, overall, decisions regarding the steps to complete the ML tasks, i.e., whether to replace the existing ML for credit lending with another one at the end of task 2, were not made until consensus was reached. By analyzing how participants navigated through joint actions between analytical phases, we found that disagreements were generally resolved through questions and elaborations. In fact, almost half of the interaction time (mean 15.5 minutes, SD 8.2, or 45\% with SD 19\%) with the visualization at the shared workspace was spent observing and discussing the visualization (i.e., passive visualization interaction). This is shown in Figure~\ref{fig:time}.  

An example of such a successful collaboration can be seen in the dialogue snippet below, where IND convinced STU to replace the GLM model being used for credit lending with another model. In fact, by interacting with the visualization, IND correctly (1) identified that the GLM systematically discriminated older customers, and (2) assessed that the provided GLM documentation was lacking relevant information (a source of risk for the management of the bank):
\begin{dialogue}
\small
\speak{STU} \emph{I think it is quite clear, what do ... what do you think?}
\speak{IND} [...] \emph{Ehm ... phew. For me the question is not really clear} [reads aloud e-mail from line manager] \emph{I mean this is not really transparent} [the GLM documentation], \emph{no, what we have here? I mean just because you are old ...} [pattern of discrimination is shown in the visualization]
\speak{STU} \emph{Yeah, okay, that is true. Yeah we don't know why it really ... that's true.}
\speak{IND} \emph{I mean there is no explanation why ...}
\speak{STU} \emph{No it only says that ehm ... these people were yes. It gives no real explanation, that's true.} 
\speak{IND} \emph{So I would say we replace it} [the GLM, as it discriminates older customers and it lacks documentation]\emph{, no?}
\speak{STU} \emph{Yeah ... yeah.}
\end{dialogue}
\normalsize

\vspace{0.1cm}
\noindent\textbf{Collaboration: Taskwork assessment.} 
The assessment of the task work performance across tasks 2, 4, and 5 (see Table~\ref{tab:performance_table} for aggregated results and the supplemental materials for all questions per topic) showed that the performance was highest at the end of task~2 when participants jointly answered questions on the 3D scatterplot of ML model predictions. In tasks 4 and 5, STU slightly outperformed IND at identifying the types of classification errors likely to be made by the GLM and the purpose of counterfactual explanations. However, IND were better at identifying data patterns, e.g., recognizing that older customers were denied the requested loans by the biased GLM in use at the bank. With respect to the confidence of participants into the ML models and explanations, we found both IND and STU to be overly confident, with STU slightly less so than IND (see Table~\ref{tab:trust_table}). In task 2, the accuracy of the model remained the same as the original model, yet participants indicated to be moderately confident in the model's superior performance, compared to the original ML model. Similarly, in task 5, all participants indicated high levels of confidence in the counterfactual explanations as global, as well as local methods, whereas the explanations are valid on a local level only.

\begin{table}[h!]
\caption{Taskwork assessment: results}
\label{tab:performance_table}
\begin{tabular}{ p{1cm} p{8.5cm} cc }
 \toprule
 \emph{Task} & \emph{Topic} & &\\ 
\hline
  &  &  \multicolumn{2}{c}{ \emph{STU and IND}}\\ 
 \hline
 Task 2 & Model type / parameter identification & \multicolumn{2}{c}{90\%}\\  
 &  Identification of correct pattern in data &  \multicolumn{2}{c}{100\%}\\
 & Decision to keep or replace ML model & \multicolumn{2}{c}{90\%} \\
  \hline
  & & \emph{IND}  & \emph{STU}\\
\hline
 Task 4 & Type of classification error the GLM is less likely to commit & 44\% & 56\%\\
        & Identification of the correct pattern in data  & 89\% & 56\%\\
 Task 5 & Purpose of counterfactual explanations & 44\% & 67\%   \\
 \bottomrule
 \multicolumn{4}{l}{\footnotesize Stated: \% correct. Joint answers by both STU and IND for task 2; separate answers for tasks 4 and 5}
\end{tabular}
\end{table}

\begin{table}[h!]
\caption{Confidence in ML model and explanations (Likert scale: 1--7)}
\label{tab:trust_table}
\begin{tabular}{ll cc}
 \toprule
 \emph{Task} & \emph{Topic} & \emph{IND}  & \emph{STU}\\
 \midrule
 Task 4 & Confidence in GLM performance based on accuracy &  5.4 (1.2) & 4.8 (1.2)\\  
 Task 5 & Confidence in counterfactuals as global explanation & 5.2 (1.1) & 4.7 (0.9)\\
        & Confidence in counterfactuals as local explanation  & 5.7 (1.4) & 5.4 (0.5)\\
 \bottomrule
 \multicolumn{4}{l}{\footnotesize Stated: mean (standard deviation)}
\end{tabular}
\end{table}

\vspace{0.1cm}
\noindent\textbf{User experience and usability.}
In the following, we provide a summary of the main results regarding the user experience and usability of our system. More details can be found in the supplemental materials. 

Overall, participants rated the system usability as ``\emph{OK}'' with a mean of 59.8 (SD 23.8) out of 100. The results of the NASA TLX questionnaire show that IND and STU both reported medium levels of mental, physical, and temporal demand. However, STU gave better ratings than IND across all items, with the exception of temporal demand. Both IND and STU positively perceived counterfactual explanations (67\% and 56\%, respectively), the visualization of the 3D scatterplot (78\% and 78\%, respectively), and the interactions with the other study participant (78\% and 56\%, respectively). However, only 33\% of STU participants perceived the interactions (e.g., air tapping) with the visualizations positively, as opposed to 78\% of IND participants. Only 44\% of IND and STU participants rated the display quality positively or neutral. The field of view of the HMD devices was perceived as positive by around half of the user group (44\% and 44\% of IND and STU, respectively). The user feedback showed that ten participants would suggest an improvement in the interactions offered by the HMDs. They suggested improving the sensitivity of air tapping, introducing other gestures, or shifting selected interactions to keyboard and mouse. 

Finally, the results demonstrate the applicability of our system. Most (67\%) of the participants rated the system positively with regard to its generalizability, as well as the possibility to use the system in order to solve real-world ML problems.

\section{Discussion}

To structure the discussion of our results, we broadly group our findings into three criteria of successful collaboration \cite{Gutwin2000TheMO}: effectiveness, efficiency, and satisfaction. Effectiveness refers to factors that contributed to or hindered the successful completion of, in our case, the ML tasks. Efficiency refers to factors that promoted or hindered the collaboration process in our CIA system. Finally, satisfaction refers to factors that contributed to or hampered participants' satisfaction and engagement with the activities and outcomes of the experience with our CIA system.

\subsection{Effectiveness}

Overall, we observed that participants quickly understood the design of our CIA system for ML modeling and were able to interact with all system components. This is supported by the usability ratings, which are in an acceptable range. Our results further show that the system design and the role-playing storyline have elicited sustained collaboration among users.

Although task division was to a certain extent dictated by the role-playing script in the study, most of the participants navigated tasks jointly and fluidly, by openly talking about their individual instructions or their current actions. This helped reveal their mental models without intervention of the experimenters, as a result of using the pair analytics methodology. Furthermore, all pairs of participants successfully completed the study experience through collaboration. The observation that one participant takes on the ``navigator'' role (usually the IND), while the other takes on the ``driver'' role (usually the STU), as supported by Figure~\ref{fig:time}, echoes related work in pair programming on the typical role dynamics among collaborators  \cite{AriasHernandez2011PairAnalytics, Chong2007TheSD, Jung2012GroupHB}. This provides evidence for the correct implementation of the pair analytics roles of SME and VAE in our study. 

\vspace{0.1cm}

\noindent\textbf{Takeaway 1: }\emph{Pair analytics can be an effective method to elicit collaboration for ML tasks in co-located, synchronous immersive settings, as it is intuitive and allows users to share context and visualizations effectively.}

\vspace{0.1cm}
\noindent The quantitative assessment of taskwork shows that most of the participants answered task 2 questions correctly but encountered difficulties in answering those from tasks 4 and 5. Although these questions were answered individually, they summarized the understanding of the collaborative activities performed by participants in tasks 4 and 5. Overall, IND performed worse than STU in answering the questions for task 4 and 5, with the exception of the identification of a pattern (i.e., older customers systematically being denied a loan) in data. We argue that industry experience, especially in data analytics, may have facilitated IND to better answer this question. Most STU (67\%) correctly answered the question on the purpose of counterfactual explanations. This may be the case, as only 3 out of 9 STU declared to have never heard of the ML interpretability method before. Overall, these results suggest that the tasks of our study may have been rather demanding for a one-time experience. Although we expected IND to perform worse than STU, we argue that the data science background of STU (not necessarily focused on ML modeling and interpretability) may have been too generic for the proposed tasks.

We further noticed lower taskwork performance in sessions where participants reached less clarity on role division and the allocation of tasks. This result shows that, also in the case of our CIA system, the division of taskwork based on the skills of collaborators is a key mechanics of successful collaboration, as remarked in  collaborative visual analytics studies \cite{Heer2008DesignCF}. 

\vspace{0.1cm}
\noindent\textbf{Takeaway 2: }\emph{Clear role division promotes effectiveness in solving ML tasks. }

\vspace{0.1cm}
\noindent Furthermore, our results on confidence indicate that IND and STU were overly confident in the model and counterfactual explanations. This may indicate that interactive visualizations of ML interpretability methods can be misleading for users, even in the case of ML-savvy ones. This finding is in line with prior work on interpretable ML using traditional displays \cite{Kaur2020InterpretingIU}.

\vspace{0.1cm}
\noindent\textbf{Takeaway 3: }\emph{Designers of CIA systems should explore ways to encourage critical thinking in order to avoid overconfidence in ML models.}

\subsection{Efficiency}

Prior work has shown that the use of multiple devices can interfere with collaborative efforts, as a shift in user focus and communication may inhibit awareness \cite{Scott2015LocalRC}. In our ML scenario, the coordination or communication between group members was not impeded by the use of HMDs and personal workspaces, indicating that our system design did not inhibit situational awareness among group members. Furthermore, due to the COVID-19 measures, all participants were requested to wear masks underneath their HMDs, covering their faces almost entirely. Our observations and the feedback we received from participants did not indicate any notable impediment to their communication and the ability to establish common ground throughout the study. 

\vspace{0.1cm}
\noindent\textbf{Takeaway 4: }\emph{The use of multiple types of interfaces (2D and 3D) does not impede collaborative efficiency when solving ML tasks.}

\vspace{0.1cm}
We noticed that participants took turns and engaged in a ``handover'' process while collaborating at the shared workspace, which hindered efficiency, as often only one person at a time was manipulating the visualization (as seen in Figure~\ref{fig:time}). Research in the CSCW literature has revealed that handover (or hand-off) and the establishment of personal spaces are in fact a critical mechanics of collaboration \cite{Pinelle2003TaskAF, Doucette2013SometimesWW}. Researchers also argue that collaborative tools should allow participants to coordinate actions and positions to avoid conflicts with those of others \cite{Gutwin2000TheMO}. Our results suggest that handovers (1) may emerge also in the case of synchronous (co-located) CIA systems, despite the initial provision of clear remarks on the synchronicity of visualizations to the users, and (2) it is a dynamics of successful collaborations.  

\vspace{0.1cm}

\noindent\textbf{Takeaway 5: }\emph{Handovers, although decreasing efficiency, are a driver of collaboration in co-located, synchronous IA; designs of CIA systems for ML should anticipate their use.}

\vspace{0.1cm}
\noindent Our results further show that participants established common ground with open communication and deixis. As remarked in previous studies \cite{Wang2019HowDS, Butscher2018ClustersTA, Moghaddam2015ProcidBC}, deixis and the ability to visually support discussions are important mechanisms of collaboration, also in the case of our CIA study. In fact, participants continuously monitored each other's attention through pointing gestures at the visualization (as seen in Table~\ref{tab:pointingTable}), or by using the visualization to support their elaborations and the establishment of common ground and consensus.

In contrast, long pauses in joint actions and silent, single-user interactions characterized a lack of common ground among participants. This appeared to be more prominent in sessions, where participants' levels of ML knowledge were lower. Collaboration was also interrupted whenever technical or conceptual challenges emerged among participants. Examples are sustained difficulties with air tapping, or with the solution of the proposed tasks. In these cases, participants began to interact with the visualization simultaneously, silently or accompanying the interactions with ``self-talk'' until a resolution of the impasse was found through the reactivation of the collaboration.

\vspace{0.1cm}
\noindent\textbf{Takeaway 6: }\emph{Designing CIA systems should consider mechanisms to (re-)establish common ground, especially for complex ML visualization (e.g., through affordances).}

\vspace{0.1cm}
\noindent The level of ML modeling knowledge of participants affected both taskwork performance and teamwork dynamics. Pairs of participants with lower levels of affinity with ML performed the worst along both dimensions of collaboration. However, participants, such as industry professionals in software or mechanical engineering positions, performed better, also in the case of confidence measures. 

\vspace{0.1cm}
\noindent\textbf{Takeaway 7:} \emph{Consensus in collaboration and effectiveness in ML modeling within CIA systems is driven by prior ML knowledge, thus highlighting the need for ML training among both user pairs.}

\subsection{Satisfaction}

Participants' user experience ratings are in an acceptable range. This, together with the findings from the user feedback questions, suggests that most of the participants perceived the visualizations provided by our proposed CIA system positively. However, as mentioned previously, participants sometimes struggled on a technical level (e.g., with air tapping) when interacting with the visualization, despite the short training at the beginning of the experience. These challenges sometimes led to visible frustration and interrupted the collaboration, as confirmed also by the analysis of the video recordings of the experimental sessions. It is likely that this affected the usability score, leading to only moderate ratings. Considering additional interactions with HMDs as an alternative to air tapping may support user experience and usability. On average, the workload was medium to low. Participants reported medium mental demand and effort, which is to be expected with the rather technical tasks they were given. The lack of extreme values is in line with the observations of the recordings, where most of the participants appeared engaged in the tasks, but did not suffer cognitive overload. 

Prior work has postulated that greater immersion leads to a greater sense of presence, in turn leading to increased engagement \cite{Bschel2018InteractionFI}, which is necessary for collaborative analytics tasks \cite{Billinghurst2018CollaborativeIA}. However, we found that the switching between 2D to immersive 3D interfaces, and vice-versa, did not have a negative effect on participants' engagement levels. Moreover, participants with higher levels of ML knowledge appeared to engage in and enjoy the collaborative activities more. We found no evidence that moving back and forth between the desktop computer at the personal workspace and the visualization (either at the personal or shared workspace) hindered or interrupted collaboration. We argue that the use of immersive interfaces along with traditional input tools, such as a keyboard and mouse, may have supported the realism of the study, benefiting the assessment of our IA system applicability to real-world scenarios, allowing most of the users to avoid levels of mental and physical overload they may have possibly incurred in the case of a system with immersive interfaces, only.

\vspace{0.1cm}
\noindent\textbf{Takeaway 8: }\emph{The use of several different interfaces (2D and 3D) does not reduce users' levels of engagement in solving ML tasks, especially for users who are more knowledgeable in ML, and may help avoid mental and physical overload.}

\subsection{Limitations}

Our work is subject to limitations. First, while in line with prior works, our user study involved only 18 participants with heterogeneous backgrounds. Second, it is needless to say that collaboration in a laboratory setting might differ from that in real-world settings. Moreover, collaboration is a social process that depends on the pairing of participants, the tasks to be solved, and the roles and power relationships among individuals.  Although we recruited a heterogeneous sample of participants with different levels of ML knowledge, the task definitions and the predetermined roles of the study participants may have biased their interactions throughout the user study. 

\section{Conclusion}

This is the first study to examine how collaboration unfolds when users with different professional backgrounds and levels of ML knowledge interact with a CIA system to solve ML tasks. Collaboration in ML modeling is characterized by several challenges, including the complex nature of ML tasks and the interdisciplinary approach in real-world ML applications where domain experts and ML experts with different skills have to collaborate. We used a methodology from collaborative visual analytics, namely pair analytics, to assess collaborative dynamics both qualitatively and quantitatively. Our results show that our proposed CIA system can elicit sustained collaboration among users during ML modeling. Our findings provide recommendations for the design of CIA systems that enable interdisciplinary teams to jointly solve ML tasks. As such, we expect our findings to be especially relevant for interdisciplinary teams in real-world ML applications from industry, and may help reduce the complexity of ML modeling among the broader workforce.



\newpage
\appendix

\end{document}